# Quasiparticle Levels at Large Interface Systems from Many-body Perturbation Theory: the XAF-GW method

Fengyuan Xuan[1], Yifeng Chen[1] and Su Ying Quek[1,2,*]

[1]Centre for Advanced 2D Materials, National University of Singapore, Block S14, Level 6, 6 Science Drive 2, 117546, Singapore

[2]Department of Physics, National University of Singapore, 2 Science Drive 3, 117542, Singapore

* To whom correspondence should be addressed: phyqsy@nus.edu.sg

**Abstract**

We present a fully *ab initio* approach based on many-body perturbation theory in the GW approximation, to compute the quasiparticle levels of large interface systems without significant covalent interactions between the different components of the interface (meaning that the different components can be separated without the creation of dangling bonds). The only assumption in our approach is that the polarizability matrix (chi) of the interface can be given by the sum of the polarizability matrices of individual components of the interface. We show analytically, using a two-state hybridized model, that this assumption is valid even in the presence of interface hybridization to form bonding and anti-bonding states, up to first order in the overlap matrix elements involved in the hybridization. We validate our approach by showing that the band structure obtained in our method is almost identical to that obtained using a regular GW calculation for bilayer black phosphorus, where interlayer hybridization is significant. Significant savings in computational time and memory are obtained by computing chi only for the smallest sub-unit cell of each component, and expanding (unfolding) the chi matrix to that in the unit cell of the interface. To treat interface hybridization, the full wavefunctions of the interface are used in computing the self-energy. We thus call the method XAF-GW (X: eXpand-chi, A: Add-chi, F: Full wavefunctions). Compared to GW-embedding type approaches in the literature, the XAF-GW approach is not limited to specific screening environments or to non-hybridized interface systems. XAF-GW can also be applied to systems with different dimensionalities, as well as to Moire superlattices such as in twisted bilayers. We illustrate the generality and usefulness of our approach by applying it to self-assembled PTCDA monolayers on Au(111) and Ag(111), and PTCDA monolayers on graphite-supported monolayer $WSe_2$, where good agreement with experiment is obtained. In this version of the paper, we provide Supplementary Material with a more detailed proof of the equations in the main text.

# 1. INTRODUCTION

The alignment of quasiparticle levels at interfaces is a long-standing and important problem. Contact resistance in electronic devices is determined by Schottky barrier heights, which are in turn determined by the energy level alignment (ELA) at the contact. The Schottky barrier height is found to be sensitive to the detailed atomic structure at the interface,[1] underscoring the importance of first principles predictions of the ELA at these interfaces. Similarly, the ELA at organic/metal interfaces determines the energy barrier for charge transfer across the interface, and is central to the functionality of organic electronic devices. The recent surge in interest in twisted bilayer two-dimensional (2D) materials,[2-4] Moire superlattices,[5, 6] and organic-2D material interfaces[7, 8] further puts into the spotlight interface systems with large unit cell sizes, where the quasiparticle band structure and ELA play a key role. It is therefore timely and highly desirable to have an *ab initio* approach to predict the quasiparticle band structures at these large interface systems.

Kohn-Sham density functional theory (DFT) calculations are the workhorse of first principles calculations, and can be used to predict the ground state atomic structure of large interface systems, including the effect of van der Waals interactions.[9, 10] However, while the DFT eigenfunctions are typically good approximations to the actual quasiparticle wavefunctions,[11] there is no theorem that guarantees that the eigenvalues of the Kohn-Sham DFT Hamiltonian are quantitatively accurate representations of the quasiparticle levels in the system (except for the eigenvalue of the highest occupied state relative to vacuum).[12] Standard local and semi-local exchange-correlation functionals result in an underestimation of the band gap. This error is particularly severe for molecules where the gap between the highest occupied molecular orbital (HOMO) and lowest unoccupied molecular orbital (LUMO) can be underestimated by several eV's, leading to an order of magnitude overestimation of electronic conductance in single molecule junctions.[13] Self-interaction corrected DFT[14] and the use of hybrid functionals incorporating a fraction of Hartree-Fock exchange alleviate this error and result in a larger quasiparticle gap. However, neither of these methods takes into account the effect of non-local screening from the environment, which can change the quasiparticle gaps significantly not only for molecules (where the gap renormalization can be ~1 eV),[15-18] but also for 2D materials on substrates.[19-21] To take into account the effect of non-local screening, a DFT+Σ approach had been proposed for molecules physisorbed on bulk substrates.[13, 15, 16, 22] The DFT+Σ approach is limited to interface systems where the molecule interacts weakly with the substrate with negligible charge transfer, and neglects the polarizability of the molecule. The method also assumes that the change in screened Coulomb potential at the position of the molecule, due to the substrate, can be approximated by an image charge potential.[15] Given the recent interest in organic/2D material interfaces, we have adapted the approach for organic/2D material interfaces, where the substrate screening potential cannot be described by a model image charge potential. Instead, we use the screening potential obtained by computing the static dielectric matrix of the 2D material in the random phase approximation (RPA), and show that the DFT+ Σ approach used in this manner can give a good estimate of the HOMO-LUMO gaps of benzene on 2D material substrates.[23] However, this approach fails to give the correct HOMO-LUMO gaps for larger molecules on 2D material substrates, where polarizability of the molecule cannot be neglected. In recent years, self-consistent DFT calculations with an optimally-tuned range-separated hybrid functional has been used to take into account the effect of screening for molecules on a metallic substrate.[24] The predicted ELA is in reasonable agreement with experiment for a range of molecule/metal substrates. However, this approach is similar in spirit to the DFT+ Σ approach, where the parameters of the functional are tuned to mimic polarization from an image potential, at the position of the molecule.[24] The evaluation of Hartree-Fock exchange in these large systems is also computationally demanding. We also remark that GW calculations for molecules that are weakly chemisorbed on metal surfaces reveal that the self-energy correction is coverage dependent - the GW HOMO level computed for a monolayer of benzene-diamine on Au(111) compares well with photoemission results,[17] but is closer to the Fermi level than that obtained for single-molecule junctions.[17, 18]

Compared to Kohn-Sham DFT, a more rigorous formalism for quasiparticle energies is many-body perturbation theory, where the electron self-energy is typically computed within the GW approximation.[11, 25] The GW approximation has been shown to give quantitatively accurate quasiparticle gaps for a large number of molecules,[26] as well as accurate quasiparticle band structures for semiconductors[11] and metals.[27] However, thus far, GW calculations have not been performed on interface systems with large unit cells, containing thousands of atoms. One major bottleneck is the computation of the dielectric matrix, or equivalently, the polarizability (chi) matrix, typically performed in many GW implementations. The computational cost for this calculation scales as $O(N^4)$ or $O(N^3 \log N)$, where N is the number of electrons.[28] Another difficulty is the need to include many unoccupied states in typical GW calculations. Recent schemes have addressed the latter difficulty with some success, such as approximating the unoccupied states with free electron states,[29] eliminating unoccupied states using the Sternheimer approximation,[30] using the so-called static remainder approach,[31] and even by achieving highly efficient parallelization of the GW code.[28] On the other hand, the requirement for large k-point meshes in calculations on some 2D material systems has also been overcome with analytical expressions for the small q limit of the screening potential[32], as well as recently developed non-uniform k-point sampling methods.[33] However, in practice, it has been found that the large memory requirement required for computing chi matrices of large interface systems, containing not only many electrons but also many reciprocal lattice vectors, severely limits the use of GW for such systems. To date, some of the largest-scale GW calculations involve liquid water and its interface with other materials - systems that contain 1000-2000 electrons per unit cell.[34, 35]

In recent years, several groups have developed embedding type methods similar in spirit to DFT+Σ,[13, 15, 16, 22] but using GW calculations. The GW calculation is performed for a particular component of the system, and electrostatic screening from the remaining components are included using either model[36, 37] or *ab initio* based approximations.[19, 21, 38, 39] For example, a hybrid GW/classical electrostatic approach has been applied to study charged excitations in molecular solids, where molecules described at the GW level are embedded in an environment modeled using classical electrostatics.[36] The effect of screening from metal substrates has also been included through an image potential in the GW calculations.[37] In other work focusing on 2D materials physisorbed on substrates, GW calculations are performed on the 2D materials, while the polarizability is modified using information from a quantum electrostatic heterostructure model,[38, 39] or by adding on the polarizability matrix of the substrate in the same supercell as the 2D material.[19, 21] Because the Add-chi method does not include explicitly the substrate wavefunctions, the method is designed for physisorbed systems only. Indeed, it is often assumed that approximating the polarizability matrix by the sum of the polarizability matrices of individual components (called the "diagonal" approximation) works only for systems where wavefunctions are localized on one component of the interface. However, in this work, we show analytically that this presumption is not true.

We propose a simple but powerful approach, called XAF-GW, which can overcome the bottleneck of the memory and computational requirement for the polarizability (chi) matrix for interface systems with large supercells and a large number of electrons per cell. Our method applies to interface systems without significant covalent interactions across the interface (meaning that the interface can be separated into different components without the creation of dangling bonds). XAF-GW is easy to implement and consists of three steps. In the "X" step, the chi matrix of the individual components of the interface is evaluated in the smallest possible sub-unit cell, and then expanded (unfolded) to a chi matrix of the same system in the unit cell for the interface. In the "A" step, the expanded chi matrices of the individual components are added together. In the "F" step, the self-energy correction is computed using the full wavefunctions of the interface and the computed chi matrix of the interface. The "X" step can be done exactly and results in huge computational savings, allowing one to perform GW calculations on large interface systems not possible before. The "F" step can be replaced by a calculation involving only wavefunctions of the component of the interface of interest, if there is no hybridization between the components. However, using the wavefunctions of the full system greatly expands the applicability of our approach in contrast to GW embedding type approaches[19, 21, 36-39] discussed above. In particular, unlike the embedding approaches, the "F" step takes into account the effect of wavefunction hybridization and non-local exchange interactions between different parts of the heterostructure, and enables the treatment of both static and dynamical screening.

In the following, we will first describe our methodology, focusing mainly on the "A" step of XAF-GW, which constitutes our only approximation in the approach. We show that the "A" step is applicable not only to physisorbed interfaces, but also to cases where the wavefunction of the heterostructure is a linear combination of wavefunctions of both components. We further show using a two-state hybridized model, that the "A" step is also applicable when bonding and antibonding states are formed between the two components, up to first order in the overlap matrix elements involved in the hybridization. Next, we provide numerical evidence for these arguments by comparing the results of XAF-GW with regular, full, GW calculations on graphene/hexagonal boron nitride (h-BN), bilayer phosphorene (where bonding and antibonding states form between the two phosphorene layers) and benzene on MoS$_2$. Finally, we apply the XAF-GW calculations to large heterostructure systems involving PTCDA on a graphite-supported WSe$_2$ substrate, PTCDA on Au(111) and PTCDA on Ag(111). These systems are chosen to illustrate the range of applicability of XAF-GW, including respectively, complex heterostructure substrates (not limited to simple metallic screening), physisorbed systems as well as more strongly chemisorbed systems. It should be emphasized that our method is fully *ab initio* and not limited to particular materials or components of the interface. In particular, applications to twisted bilayer systems are currently underway.

## 2. METHODOLOGY

The XAF-GW approach can be applied to interface systems without significant covalent interactions. This is because the approach requires a partitioning of the interface into individual components, which should not have dangling bonds. Examples where the XAF-GW approach would not work include thiolate molecules on Au(111), where the S-H bond in the molecule breaks to form a covalent bond between the molecular radical and the Au substrate. However, XAF-GW is applicable to a wide variety of interfaces, including the plethora of "van der Waals heterostructure" systems, organic/metal interfaces, organic/2D material interfaces, as well as more strongly interacting interfaces where some degree of hybridization is present, such as in black phosphorus[40-42] and other layered materials.[43] Following the identification of the separate components in the heterostructure, XAF-GW consists of three steps: eXpand-chi, Add-chi, and Full-Sigma. In the eXpand-chi ("X") step, the smallest sub-unit cell for each separate component is identified, and the chi matrix of the component is computed using this sub-unit cell, after which the chi matrix of the same component in the supercell of the heterostructure is obtained by an expansion (unfolding) procedure, described in Section 2.3. The reciprocal space sampling used in these calculations should be compatible with the target reciprocal space sampling to be used in the supercell. Since the computation of the chi matrix is the bottleneck in GW calculations, the "X" step provides substantial computational savings, reducing the computational cost of the calculation by ~$O(m^6)$, and the memory requirement by ~$O(m^4)$. In the Add-chi ("A") step, the expanded chi matrices from all components of the interface are added together, and in the Full-sigma ("F") step, the self-energy corrections are computed using the full set of wavefunctions for the heterostructure, and the approximated chi matrix.

## 2.1. GW method as implemented in the BerkeleyGW code

Our XAF-GW code is based on the BerkeleyGW[28] code, which demonstrates good parallelization efficiency. In this sub-section, we provide some background on the BerkeleyGW implementation of the GW method. However, we emphasize that the approach described in this work is general, and is not limited to specific GW implementations.

The GW method is a many-body perturbation theory approach to compute the quasiparticle levels of any given system, given only its atomic structure. The starting point of the calculation is a mean-field calculation, typically performed using Kohn-Sham DFT with standard exchange-correlation functionals. We have:

$$\left[-\frac{\nabla^2}{2}+V_{ion}+V_H+V_{xc}^{DFT}\right]\psi_{n\vec{k}}^{DFT}=E_{n\vec{k}}^{DFT}\psi_{n\vec{k}}^{DFT} \qquad (1)$$

where $E_{n\vec{k}}^{DFT}$ and $\psi_{n\vec{k}}^{DFT}$ are the Kohn-Sham eigenvalues and eigenfunctions, respectively. These Kohn-Sham eigenvalues and eigenfunctions are used as a starting guess for the quasiparticle eigenvalues and eigenfunctions, respectively. The latter are computed by solving the following equation:

$$\left[-\frac{\nabla^2}{2}+V_{ion}+V_H+\Sigma\left(E_{n\vec{k}}^{QP}\right)\right]\psi_{n\vec{k}}^{QP}=E_{n\vec{k}}^{QP}\psi_{n\vec{k}}^{QP} \qquad (2)$$

where $E_{n\vec{k}}^{QP}$ and $\psi_{n\vec{k}}^{QP}$ are the Kohn-Sham eigenvalues and eigenfunctions, respectively, and $\Sigma$ is the self-energy operator within the GW approximation, i.e., $\Sigma = i$GW, where G refers to the one-particle Green's function, and W refers to the dynamically screened Coulomb interaction.

After obtaining $E_{n\vec{k}}^{DFT}$ and $\psi_{n\vec{k}}^{DFT}$, the first step of the calculation is the computation of W. We compute explicitly the static RPA polarizability (chi) matrix, as follows:

$$\chi_{\vec{G}\vec{G}'}(\vec{q};0) = \sum_v \sum_c \sum_{\vec{k}} M_{vc}(\vec{k},\vec{q},\vec{G}) M_{vc}^*(\vec{k},\vec{q},\vec{G}') \frac{1}{E_{v,\vec{k}+\vec{q}} - E_{c,\vec{k}}} \quad (3)$$

where the matrix elements

$$M_{vc}(\vec{k},\vec{q},\vec{G}) = \langle v, \vec{k}+\vec{q} | e^{i(\vec{q}+\vec{G})\cdot\vec{r}} | c, \vec{k} \rangle \quad (4)$$

Here, v refers to valence band (occupied) states, and c refers to conduction band (unoccupied) states. $\vec{q}$ is a vector in the first Brillouin zone, while $\vec{G}$ is a reciprocal lattice vector. The evaluation of Equation (3) is $O(N^3 \log N)$. The eigenvalues and wavefunctions used in Equation (3) are the mean-field Kohn-Sham eigenvalues and eigenvectors, in what is called the $G_0W_0$ approximation. The eigenvalues in Equation (3) can also be updated with the computed $G_0W_0$ quasiparticle energies, and the resulting chi matrix is used in a subsequent GW calculation, giving the $G_0W_1$ approximation. Using the above expression for chi, we can obtain the static RPA dielectric matrix as

$$\varepsilon_{\vec{G}\vec{G}'}(\vec{q};0) = \delta_{\vec{G}\vec{G}'} - v(\vec{q}+\vec{G})\chi_{\vec{G}\vec{G}'}(\vec{q};0) \quad (5)$$

where $v(\vec{q}+\vec{G})$ is the bare Coulomb interaction. As the heterostructures considered in this work are slab systems, we use a slab Coulomb truncation[44] expression for $v(\vec{q}+\vec{G})$. The static screened Coulomb interaction is then obtained using the expression:

$$W_{\vec{G}\vec{G}'}(\vec{q};0) = \varepsilon^{-1}_{\vec{G}\vec{G}'}(\vec{q};0)v(\vec{q}+\vec{G}')\tag{6}$$

The dynamically screened Coulomb interaction is obtained using the Hybertsen-Louie generalized plasmon pole (GPP) approximation.[11]

Instead of the GPP approximation, the full frequency dependence of the dielectric matrix can be computed explicitly using the frequency-dependent expression for the chi matrix:

$$\chi_{\vec{G}\vec{G}'}(\vec{q};\omega) = \sum_v \sum_c \sum_{\vec{k}} M_{vc}(\vec{k},\vec{q},\vec{G})M^*_{vc}(\vec{k},\vec{q},\vec{G}')\frac{1}{\omega+E_{v,\vec{k}+\vec{q}}-E_{c,\vec{k}}}\tag{7}$$

Once the screened Coulomb interaction is obtained using the chi matrix, the self-energy (sigma) is then computed, using formula obtained from the expression, $\Sigma = i$GW. The Green's function, G, is given by

$$G(\vec{r},\vec{r}';E) = \sum_{n,\vec{k}} \frac{\phi_{n\vec{k}}(\vec{r})\phi^*_{n\vec{k}}(\vec{r}')}{E-\varepsilon_{n\vec{k}}-i\delta_{n\vec{k}}}\tag{8}$$

where $\phi_{n\vec{k}}(\vec{r})$ is approximated by the Kohn-Sham eigenfunction, and $\delta_{n\vec{k}} = 0^+$ for occupied states and $\delta_{n\vec{k}} = 0^-$ for unoccupied states. In the $G_0W_0$ approximation, the eigenvalues $\varepsilon_{n\vec{k}}$ in equation (8) are given by the Kohn-Sham eigenvalues. These can be updated to the $G_0W_0$ quasiparticle energies and the self-energy recomputed, in the $G_1W_0$ approximation. In the $G_1W_1$ approximation, the eigenvalues in equation (3) (or (7)) are also updated as described above.

In the Full-Sigma step of XAF-GW, the wavefunctions of the entire heterostructure are used in equation (8) for the computation of sigma. In contrast, the GW embedding approaches in the literature[19, 21, 36-39] typically use only the wavefunctions of the component of interest in equation (8).

Typically, the Kohn-Sham eigenfunctions are a good approximation to the quasiparticle wavefunctions, except in the case of strongly hybridized interfaces, such as benzene-dithiol on Au, where one part of the interface has a much larger self-energy correction. Methods have been developed to deal with such systems,[17, 45-47]. Where necessary, we will discuss the implications of these considerations on our predictions.

## 2.2. Applicability of the Add-chi approximation

The Add-chi approximation is that

$$\chi^{HS}_{\vec{G}\vec{G}'}(\vec{q};\omega) = \chi^{1}_{\vec{G}\vec{G}'}(\vec{q};\omega) + \chi^{2}_{\vec{G}\vec{G}'}(\vec{q};\omega) \tag{9}$$

We now proceed to evaluate the conditions under which the Add-chi method is appropriate, noting that this is the only approximation in the XAF-GW method. We shall focus on the static case with $\omega = 0$, because generalization to non-zero $\omega$ is trivial. Let us consider an interface that can be partitioned into two components, labeled "1" and "2", without the creation of dangling bonds. We shall evaluate three different cases.

Case 1: The wavefunctions of the heterostructure (HS) are all completely localized on 1 or 2 only, and are given by the Kohn-Sham eigenfunctions of the individual components 1 or 2. The Kohn-Sham eigenspectrum of these states is the same as those of the individual components, but may be shifted by an additive constant.

Case 2: The wavefunctions of the HS are not completely localized on 1 or 2 only, but consist of a linear combination of the Kohn-Sham eigenfunctions of 1 and 2. Specifically, we assume that when 1 and 2 come into contact, the eigenvalues of 1 and 2 may shift relative to one another, but only by an additive constant (keeping the same energy differences between eigenvalues). When the eigenvalues associated with 1 and 2 are degenerate, the wavefunctions of the HS can be a linear combination of the corresponding eigenfunctions of 1 and 2. However, we assume that the eigenfunctions of 1 and 2 form an orthonormal set (in particular, eigenfunctions of 1 and 2 are assumed not to overlap in this case).

Case 3: We extend our considerations in Case 2 by allowing for hybridization between 1 and 2, with non-zero overlap between eigenfunctions of 1 and 2. We define a two-state hybridized model as follows. The valence band states of 1 and 2 overlap to form bonding and antibonding combinations in the HS, and likewise, the conduction band states of 1 and 2 overlap to form bonding and antibonding combinations in the HS.

We shall show that the Add-chi approximation holds for the above three cases. Full derivations are provided in the Appendix. In this sub-section, we briefly explain the physics behind these derivations.

Case 1 is the simplest and most obvious case. The fact that the Add-chi approximation holds for Case 1 follows immediately from the expression for chi (equations 3 and 7).

In Case 2, we allow the wavefunctions of the HS to be linear combinations of the wavefunctions of 1 and 2. However, we do not allow overlap between wavefunctions of 1 and 2. This implies that the wavefunctions of the HS are a unitary transformation of the wavefunctions of 1 and 2. The fact that the transformation is unitary, together with the orthonormality of the wavefunctions, results in the Add-chi expression (equation (9)) without any further assumptions.

In Case 3, we first obtain analytically the expression for the eigenvalues and eigenfunctions of the HS when valence states of 1 and 2 are allowed to hybridize, and also, when conduction states of 1 and 2 are allowed to hybridize. Substituting these expressions into the expression for chi, we can show that the Add-chi expression (equation (9)) holds, up to first order in the matrix elements involved in the hybridization. In other words, the first order terms for matrix elements involved in the hybridization work out to be exactly zero.

We thus show that the Add-chi approximation works in systems where the heterostructure wavefunction is a linear combination of the wavefunctions on the individual components of the interface, as well as cases where interface hybridization occurs to form bonding and antibonding states. Numerical evidence will be provided in Section 3.1.

### 2.3. eXpand-Chi procedure

For a given real space cell, the polarizability (chi) matrix is defined in the reciprocal space cell as $\chi_{\vec{G}\vec{G}'}(\vec{q})$, where $\vec{G}$ and $\vec{G}'$ run over the reciprocal space lattice vectors up till the dielectric matrix cutoff, and $\vec{q}$ runs over the mesh points that are sampled within the first Brillouin Zone. Its relation to real space $\chi(\vec{r},\vec{r}')$ is given as:

$$\chi(\vec{r},\vec{r}') = \frac{1}{\Omega} \sum_{\vec{q},\vec{G},\vec{G}'} e^{i(\vec{q}+\vec{G}).\vec{r}} \chi_{\vec{G}\vec{G}'}(\vec{q}) e^{-i(\vec{q}+\vec{G}').\vec{r}'} \quad (10)$$

where $\Omega$ is the crystal volume.

The eXpand-chi procedure consists of mapping the indices $\vec{G}$, $\vec{G}'$ and $\vec{q}$ in the sub-unit cells to those in the actual unit cell of the heterostructure (HS), so that the real space $\chi(\vec{r},\vec{r}')$ is the same. This is possible when the $\vec{q}$ sampling in the sub-unit cell is chosen to be compatible (with the same mesh density) as that in the unit cell of the HS. Consider the case where the HS unit cell is m x m of the sub-unit cell. Then the first brillouin zone (BZ) of the sub-unit cell is m x m of the BZ of the HS unit cell. The set of $\vec{q}$ vectors, $\{\vec{q}_{HS}\}$, in the BZ of the HS unit cell is a subset of that in the BZ of the sub-unit cell $(\{\vec{q}_{sub}\})$, while the set of $\vec{G}$ vectors for the sub-unit cell $(\{\vec{G}_{sub}\})$ is a subset of that for the HS unit cell $(\{\vec{G}_{HS}\})$.

Thus, given $\chi_{\vec{G}\vec{G}'}(\vec{q})$ for the sub-unit cell, we can define $\chi_{\vec{G}\vec{G}'}(\vec{q})$ for the HS unit cell as follows. For $\vec{q}_{sub} \in \{\vec{q}_{HS}\}$, the correspondence is trivial: $\chi_{\vec{G}_{sub}\vec{G}_{sub}'}(\vec{q}_{sub}) = \chi_{\vec{G}_{HS}\vec{G}_{HS}'}(\vec{q}_{HS})$. For $\vec{q}_{sub} \notin \{\vec{q}_{HS}\}$, we can define $\vec{q}_{sub} = \vec{q}^I_{HS} + \vec{G}^I_{HS}$ for some $\vec{q}^I_{HS}$ and $\vec{G}^I_{HS}$. Then, $\chi_{\vec{G}_{sub}\vec{G}_{sub}'}(\vec{q}_{sub}) = \chi_{\vec{G}_{HS}\vec{G}_{HS}'}(\vec{q}^I_{HS})$, where $\vec{G}_{HS} = \vec{G}_{sub} + \vec{G}^I_{HS}$ and $\vec{G}'_{HS} = \vec{G}'_{sub} + \vec{G}^I_{HS}$. All other terms in $\chi_{\vec{G}_{HS}\vec{G}_{HS}'}(\vec{q}_{HS})$ are zero.

Code for the XAF-GW method is available at https://github.com/quek-group/XAF-GW.

## 2.4. Technical details

In this work, we perform calculations on graphene on monolayer (ML) hexagonal boron nitride (hBN), bilayer black phosphorus (BP), benzene on ML $MoS_2$ (3 x 3 cell), PTCDA on ML $WSe_2$ supported on bilayer graphene, PTCDA on Au(111) and PTCDA on Ag(111). Full GW calculations are performed for benchmark purposes on all systems except PTCDA on substrates. Our GW calculations are performed using the BerkeleyGW[28] package. The starting point of the GW calculations are DFT eigenvalues and wavefunctions obtained using norm conserving pseudopotentials in the plane-wave Quantum Espresso[48] code. Optimized norm-conserving pseudopotentials[49] are used for all elements except for phosphorus. These pseudopotentials are optimized to give excellent results for a kinetic energy cutoff of 60 Ry[49], which is used in this work. For phosphorus, we use a kinetic energy cutoff of 55 Ry, following our previous work.[20] In our DFT calculations, we use the PBE exchange-correlation functional[50] for all systems .To properly take into account the effect of exchange on the self-energy corrections, semi-core states are included in the pseudopotentials, giving 19 electron pseudopotentials for Au and Ag, and a 28 electron pseudopotential for W. The slabs in the calculations are separated from their periodic images with a vacuum height of at least 14 Å. The GW calculations are performed with a slab Coulomb truncation.[44]

The atomic structures of graphene/hBN, bilayer BP, and PTCDA on Ag(111) were obtained by relaxing the forces until the forces on all atoms were less than 0.001 Ry/au. These forces were computed using PBE + D2 (Grimme's dispersion correction[51]). The atomic structure of benzene on ML MoS$_2$ (3 x 3 cell) is taken from Ref. 23, while those for PTCDA on ML WSe$_2$ supported on bilayer graphene, and PTCDA on Au(111) are taken from Ref. 52 and [53], respectively.

Table I shows other computational details used in the calculations. We note that although it has been shown that very dense sampling of the reciprocal space is needed for converging the optical spectra of 2D transition metal dichalcogenides like WSe$_2$, the k-mesh chosen for our calculation results in a converged screening potential in the region of PTCDA. Doubling the k-mesh for PTCDA on WSe$_2$ also does not change the DFT band structure. Increasing the number of bands from 2200 to 2500 does not change the GW PDOS for PTCDA on Au(111) and Ag(111).

| System | k-mesh | Dielectric matrix cutoff (Ry) | Number of bands |
| --- | --- | --- | --- |
| graphene/hBN | 12 x 12 x 1 | 10 | 1000 |
| bilayer BP | 10 x 14 x 1 | 15 | 600 |
| benzene/MoS$_2$ | 6 x 6 x 1 | 16 | 2700 |
| PTCDA/WSe$_2$/graphite | 4 x 2 x 1 | 8 | 2000 |
| PTCDA/Au | 4 x 2 x 1 | 8 | 2200 |
| PTCDA/Ag | 4 x 2 x 1 | 8 | 2200 |

Table I. Computational parameters used in the GW calculations.

3. RESULTS AND DISCUSSION

### 3.1. Benchmarks with Full GW calculations

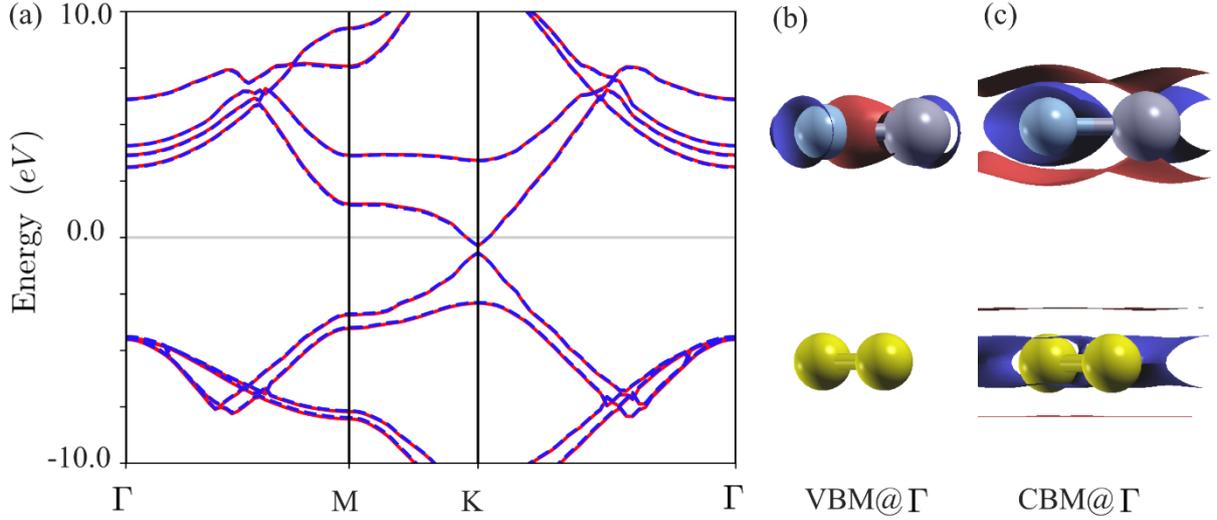

FIG. 1 Electronic structure of graphene/hBN. (a) XAF-GW (red) and full GW (blue dashed) band structures, (b) valence band maximum (VBM) wavefunction at Γ, (c) conduction band minimum (CBM) wavefunction at Γ. The isocontour value was taken to be 10 % of the maximum. Yellow, blue and gray balls represent C, N and B, respectively.

In this sub-section, all our GW calculations are done within the $G_0W_0$ approximation. Figure 1a shows the GW band structure of bernal-stacked graphene/hBN. It is clear that the XAF-GW band structure agrees very well with the full GW band structure, computed with the same convergence parameters, but without the Add-chi approximation. The graphene/hBN system is weakly interacting without significant hybridization between the two layers. The wavefunction shown in Fig. 1b (VBM @ Γ) is localized only on hBN (case 1 in Section 2.2), while that in Fig. 1c (CBM @ Γ) is a linear combination of wavefunctions on hBN and graphene (case 2 in Section 2.2), with 66% of the wavefunction localized on hBN. Following Ref. 19, we can also use only the hBN wavefunctions in the calculation of the self-energy (XA(noF)-GW). We find that XA(noF)-GW gives an error in the band gap at Γ, of ~1.3 eV (XA(noF)-GW: 6.33 eV, XAF-GW: 7.65 eV, full GW: 7.59 eV). This shows that using the wavefunctions of the full system for computing the self-energy is important when the wavefunctions of the heterostructure are modified from those in the individual components, as shown in Fig. 1c. As a consistency check, we note that at the K point, where the VBM and CBM wavefunctions of hBN are completely localized on hBN, the XA(noF)-GW gap is within 0.1 eV of that obtained with XAF-GW and full GW.

While Fig. 1 shows excellent agreement between XAF-GW and full GW band structures for the relaxed graphene/hBN interface, it is instructive to check how predictive XAF-GW is when the distance between the graphene and hBN monolayer is artificially decreased. This is because we expect that as the interlayer distance decreases, some of our underlying assumptions in our derivations in Section 2.2, such as neglecting $g_{ij}$ and $g'_{ij}$ for $i \neq j$ (see Appendix), can become questionable. Fig. 2 shows the results for the hBN band gap at the K point for different interlayer distances $d$ between graphene and hBN. For $d \sim 3.4$ Å (close to the optimal distance between graphene and hBN), the full GW, XAF-GW and XA(noF)-GW gaps agree very well. As $d$ decreases, we find, surprisingly, that the difference between the XAF-GW and full GW gaps is still quite small, although there is a notable increase in the error as $d$ decreases from 2.6 Å to 2.2 Å. At $d = 1.8$ Å, the maximum projection of the hBN VBM and CBM at K on the interface wavefunctions are, respectively, 63% and 83%. On the other hand, the XA(noF)-GW result shows a much larger increase in error as $d$ decreases, because the percentage of the wavefunctions localized on hBN decreases. We analysed the origin of the discrepancy (~2.0 eV) between XAF-GW and XA(noF)-GW gaps in detail for $d = 1.8$ Å, and found that most of the error arises at the difference in the mean-field DFT gaps obtained using the two approaches (~1.5 eV), while the remaining ~0.5 eV error arises from a difference in the self-energies computed with the two approaches. This analysis suggests that a better approximation with XA(noF)-GW may be obtained by correcting the eigenvalues at the DFT level, as suggested in the G$\Delta$W method in Ref. 38, although some errors are still expected in that case.

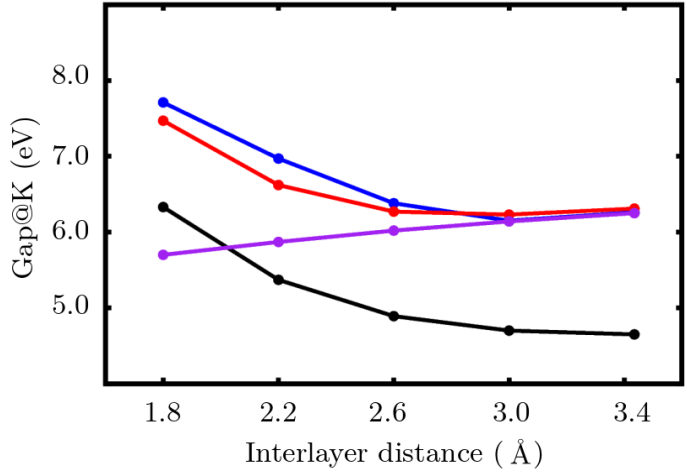

FIG. 2. hBN band gap at K point as a function of distance between graphene and hBN monolayer, computed using DFT (black), full GW (blue), XAF-GW (red) and XA(noF)-GW (purple).

The interlayer interaction in black phosphorus is significantly larger than that in graphene and hBN[41, 42, 54], giving rise to a large change in the frequency of the interlayer breathing mode with thickness,[40] and an electrical field induced Dirac cone.[55] In Fig. 3b-e, we can see clearly that the wavefunctions on individual phosphorene layers hybridize to form bonding and antibonding combinations in bilayer BP. Bilayer BP is therefore a good test system for Case 3 considered in Section 2.2. The XAF-GW band structure agrees very well with the full GW band structure for bilayer BP (Fig. 3a), thus showing that our analysis for Case 3 (see also Appendix) is valid for this system. We note that XA(noF)-GW (performed by using the wavefunctions of monolayer BP in the calculation of the self-energy) gives a band gap of 1.64 eV, ~0.3 eV larger than the XAF-GW and full GW gaps of 1.36 eV and 1.30 eV, respectively. This is consistent with the expectation that in general, larger self-energy corrections are obtained for wavefunctions that are more localized (as is the case for wavefunctions on monolayer BP compared to bilayer BP). Interestingly, even when the interlayer distance between the BP layers is decreased from 3.10 Å to 2.0 Å, the XAF-GW gap (2.02 eV) is still very close to the full GW gap (1.96 eV), while the error in the self-energy corrections for the bands at Γ are at most 0.14 eV. These results show that our assumptions in Section 2.2 seem to be valid even at these relatively small interlayer distances.

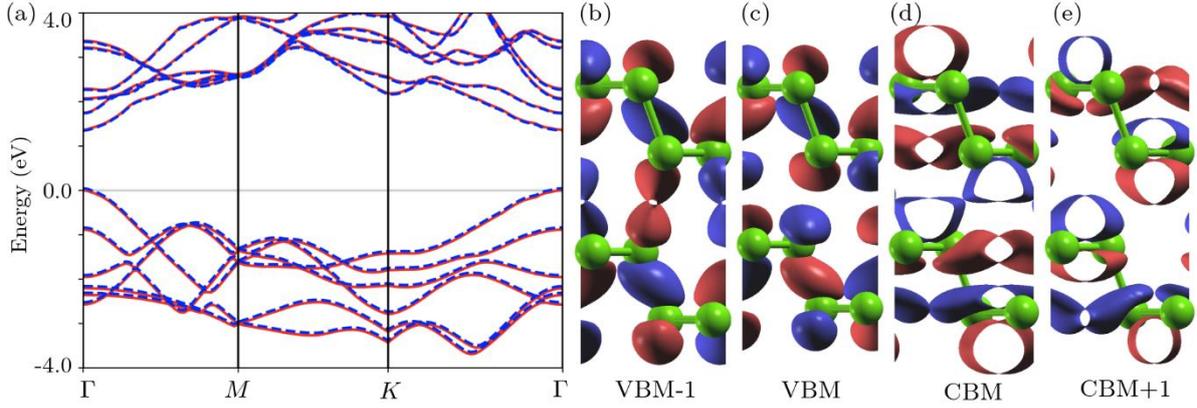

FIG. 3 Electronic structure of bilayer BP. (a) XAF-GW (red) and full GW (blue dashed) band structures, (b) (VBM-1) wavefunction at Γ, (c) valence band maximum (VBM) wavefunction at Γ, (d) conduction band minimum (CBM) wavefunction at Γ. (e) (CBM+1) wavefunction at Γ. The isocontour value was taken to be 16 % of the maximum for b and c, and 11% for d and e.

We have shown that the Add-chi approximation works well in practice for example systems corresponding to cases 1 to 3 in Section 2.2. The eXpand-chi procedure has been checked rigorously by comparing individual elements of the chi matrix with that computed explicitly for the supercell. The GW band gaps of supercells of 2D materials have also been used to further verify our expansion (unfolding) procedure. For benzene adsorbed 3.25 Å above ML $MoS_2$ in a 3 x 3 cell, XAF-GW gives a HOMO-LUMO gap of 7.82 eV, in good agreement with 7.66 eV obtained using a full GW calculation with the same computational parameters.

### 3.2. PTCDA monolayer on substrates

The energy level alignment (ELA) at organic/inorganic interfaces is a critical parameter for the functionality of organic electronic and optoelectronic devices. However, an accurate prediction of the ELA at such interfaces has remained elusive, especially for large molecules with substantial polarizability, or systems with interfacial charge transfer, where the DFT+Σ method[13, 15, 16, 22, 23] cannot be applied reliably. In this section, we apply the XAF-GW method to study the ELA of PTCDA monolayers on various substrates. These systems are chosen to illustrate the versatility of XAF-GW, which takes into account the polarizability of the molecules as well as both static and dynamical screening effects on the ELA, and does not assume an image charge potential for the substrate. In particular, the ELA of PTCDA on substrates has been studied extensively by both photoemission[56] and scanning tunneling spectroscopy (STS) experiments,[52, 57-59] and is thus an excellent benchmark system.

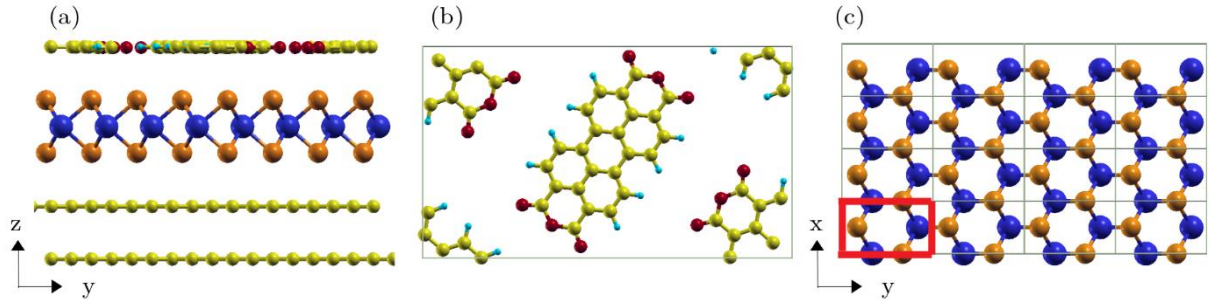

FIG. 4 PTCDA on monolayer $WSe_2$ supported on bilayer graphene. (a) Side view of atomic structure. (b) Top view showing PTCDA molecules only in one unit cell. (c) Top view showing $WSe_2$ monolayer only in one unit cell of the heterostructure. The red box indicates the sub-unit cell used for computing the chi matrix of $WSe_2$.

In all systems studied here, the PTCDA molecules self-assemble in a herringbone pattern as shown in Figure 4b. We first compute the GW HOMO-LUMO gap for this PTCDA ML, where we have converged the gap size with respect to the k-mesh. The $G_0W_0$ HOMO-LUMO gap for the PTCDA ML is 4.5 eV, while that of the gas phase PTCDA molecule is 4.8 eV. Following Ref. 60, we also perform a $G_1W_1$ calculation for these systems, and obtain gaps of 4.8 eV and 5.1 eV for the monolayer and gas phase PTCDA molecules, respectively. The latter is consistent with a ΔSCF calculation for the HOMO-LUMO gap.[52]

Figure 4 shows the atomic structure of PTCDA on monolayer $WSe_2$ supported on bilayer graphene. The sub-unit cell used for computing the chi matrix for $WSe_2$ is shown in the red box in Fig. 4c. The supercell size is 16 times that of the sub-unit cell. Given that we are using 28 electron pseudopotentials for W, the eXpand-chi method is essential to keep the calculation tractable. The HOMO-LUMO gap in the projected density of states (PDOS) on PTCDA, predicted using XAF-GW is 3.43 eV, in reasonable agreement with the experimentally measured STS gap of 3.7 eV[52] (Fig. 5). The level alignment also matches well with experiment. In this calculation, we have updated the eigenvalues in W, as well as the eigenvalues for about 60 bands near the Fermi level in the Green's function ($G_1'W_1$ calculation). We note that the HOMO wavefunction at Γ for PTCDA/$WSe_2$ is not completely localized on PTCDA. To account for the possibility that the DFT eigenfunctions may not be a good representation of the quasiparticle wavefunctions, we estimated the levels at Γ using the approach developed in Ref. 45 and 17, evaluating the self-energy in the basis of molecular orbitals. The resulting HOMO and LUMO levels at Γ are unchanged.

Table II shows the HOMO-LUMO gap at Γ for PTCDA on ML $WSe_2$, and PTCDA on ML $WSe_2$ supported on bilayer graphene. We note that the two inequivalent PTCDA molecules in the herringbone structure result in the HOMO and LUMO of the PTCDA ML being localized on different molecules. However, the HOMO-LUMO gaps of each molecule in the PTCDA ML are very similar, and their average value is reported here. This detail does not affect the HOMO-LUMO gaps obtained from the PDOS plots. The DFT PBE HOMO-LUMO gap for PTCDA on these substrates is very close to that for the gas phase molecule (1.47 eV), consistent with the absence of non-local screening in the PBE calculations. On the other hand, the DFT + Σ HOMO-LUMO gaps are larger and are substrate-dependent, but result in values that are too small. This trend is consistent with the observation that neglecting the polarizability of the adsorbates leads to gaps that are too small.[37, 52, 61] Comparing the XAF-GW HOMO-LUMO gaps @ Γ for PTCDA/WSe$_2$ and PTCDA/WSe$_2$/bilayer graphene, we see that bilayer graphene reduces the HOMO-LUMO gap by about 0.46 eV, while the WSe$_2$ ML already leads to substantial gap renormalization.

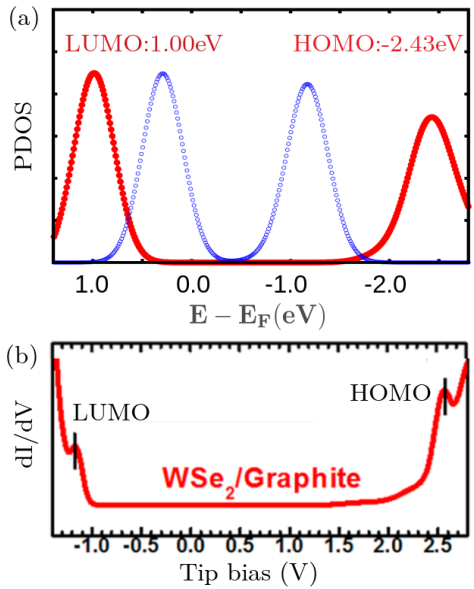

FIG. 5 (a) Computed PDOS for PTCDA on WSe$_2$/bilayer graphene. Red: XAF-GW, Blue: DFT, (b) STS spectra for PTCDA on WSe$_2$/bilayer graphene, from Ref. 52.

| System | XAF-GW | DFT PBE | DFT+Σ (image potential) | DFT+Σ (RPA screening potential) | Experiment |
|---|---|---|---|---|---|
| PTCDA/$WSe_2$ | 3.94 | 1.49 | - | 2.47 | - |
| PTCDA/$WSe_2$ on bilayer graphene | 3.48 (3.43) | 1.49 | 3.29 | 2.18 | 3.73 |

Table II. HOMO-LUMO gaps at Γ of PTCDA on $WSe_2$ ML and PTCDA on $WSe_2$ ML supported on bilayer graphene. In brackets is the HOMO-LUMO gap obtained from the PDOS plot. The XAF-GW gaps are obtained in the $G_1'W_1$ approximation. The corresponding XAF-GW gaps in the $G_0W_0$ approximation are 3.88 eV for PTCDA/$WSe_2$, and 3.57 eV for PTCDA/$WSe_2$/bilayer graphene. The values for DFT+Σ (image potential) and experiment are obtained from Ref. 52.

The ELA for PTCDA on Au(111) and Ag(111) substrates are also computed. Care is taken to ensure that the interlayer distances $d$ between the ML and the substrates are very close to the available experimental values[53, 58]. For PTCDA on Au(111), where the molecules exhibit negligible distortion on the substrate, we use the experimental lattice parameters and experimental value of $d$ (3.27 Å[53]). For PTCDA on Ag(111), the PTCDA molecules are distorted relative to their planar gas phase geometries. Therefore, the atomic structure for PTCDA on Ag(111) was obtained by first principles calculations. Using the PBE+D2 exchange-correlation functional, we obtain an interlayer distance $d$ of 2.87 Å, in excellent agreement with experiment (2.86 ± 0.01 Å[58]). In the relaxed geometry (Fig. 6d), the PTCDA molecule on Ag(111) is distorted by Δz of 0.33 Å, in good agreement with the experimental value of ~0.45 Å.[58]

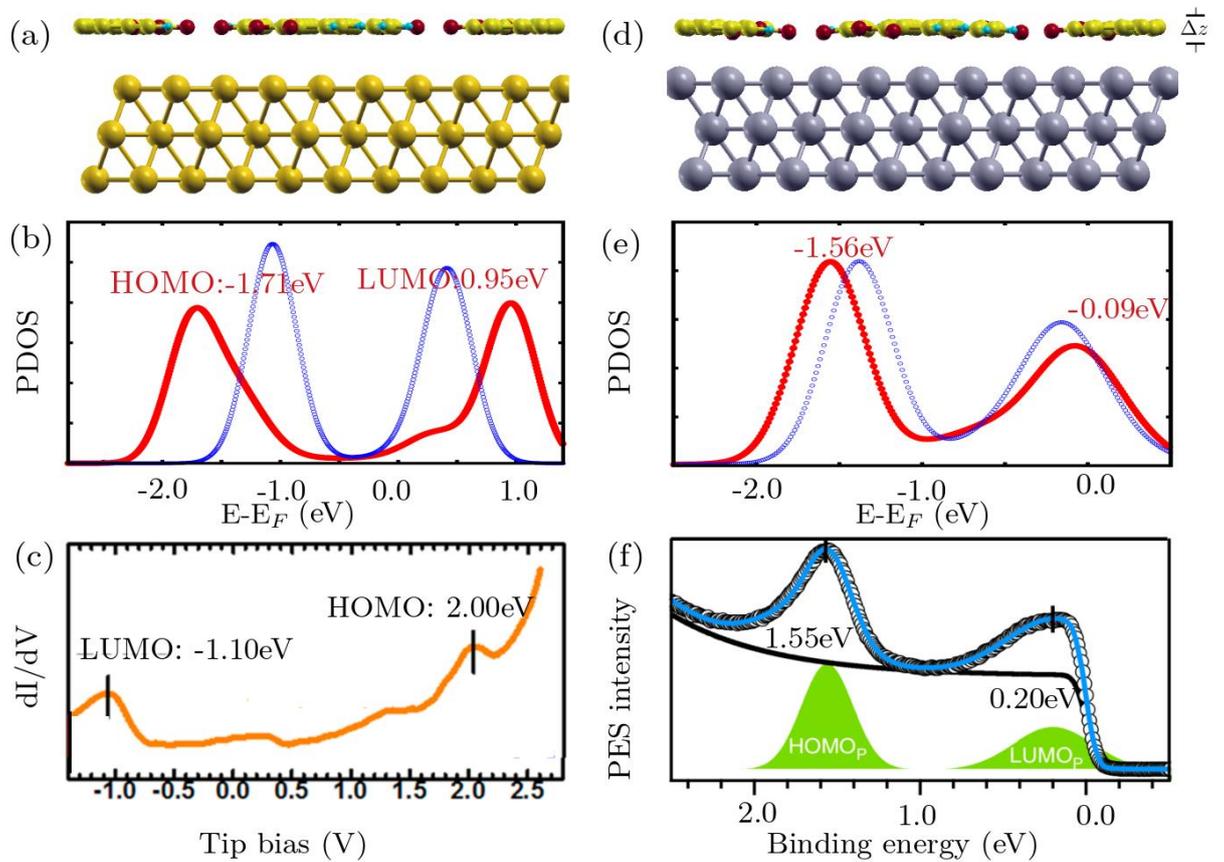

FIG. 6. Atomic and electronic structure of PTCDA on Au(111) (left) and Ag(111) (right). (a) Atomic structure of PTCDA/Au(111), (b) Computed PDOS for PTCDA on Au(111). Red: XAF-GW, Blue: DFT, (c) STS spectra for PTCDA on Au(111), from Ref. 52, (d) Atomic structure of PTCDA/Ag(111), (e) Computed PDOS for PTCDA on Ag(111). Red: XAF-GW, Blue: DFT, (f) Photoemission spectra for PTCDA on Ag(111), from Ref. 58.

From Fig. 6, we see that XAF-GW gives good quantitative agreement with the experimental ELA for PTCDA on Au(111) and Ag(111) substrates. (We note that the photoemission results shown in Fig. 6f are very similar to the STS spectra from Ref. 59) The HOMO-LUMO gap is substantially larger for PTCDA on Au(111) than for PTCDA on Ag(111), where the LUMO is nearly completely occupied. The fact that XAF-GW can reproduce these distinctive qualitative trends in experiments shows the versatility of the approach. We remark that we have also increased the number of bands in the calculation from 2200 to 2500 and found no changes to the PDOS for PTCDA on Au(111) and Ag(111). PTCDA does not interact strongly with Au(111), but does interact strongly with Ag(111). However, the fact that the DFT level alignment is so close to the GW and experimental level alignment indicates that the DFT starting wavefunctions are a good approximation to the quasiparticle wavefunctions for PTCDA on Ag(111). Accordingly, our PTCDA/Ag(111) calculations are done in the $G_0W_0$ approximation, while the PTCDA/Au(111) calculations are done in the $G_1'W_1$ approximation.

## 4. CONCLUSION

We have introduced a fully *ab initio* approach, XAF-GW, that can compute, within the GW approximation, the quasiparticle levels of large interface systems that can be partitioned into individual components without the creation of dangling bonds. In contrast to many of the embedding-like approaches in the literature that are targeted at large interface systems, our method does not require assumptions of the form of the screening potential, or assumptions about the absence of interface hybridization. Our only approximation is that the polarizability matrix of the interface system can be expressed as a sum of the polarizability matrices of the separate components of the interface (Add-chi approximation). We show, for the first time, that the Add-chi approximation holds for systems where the interface wavefunctions are a linear combination of the wavefunctions of individual components, and even for systems with some interface hybridization. The latter is proven within the context of a two-state hybridization model where the interface valence and conduction band wavefunctions are, respectively, bonding and anti-bonding combinations of the valence and conduction band wavefunctions of individual components. Within this model, the Add-chi approximation holds up to first order in the overlap matrix elements involved in the hybridization. We show numerically that this approximation holds true for bilayer black phosphorus, where the interlayer hybridization is apparent. Computational memory and time are saved by using the eXpand-chi step[62] in XAF-GW, where the polarizability matrices of individual components are computed for the smallest sub-unit cell, and then expanded to the required matrix in the interface supercell. We illustrate our approach using PTCDA monolayers on Au(111), Ag(111) and graphite-supported monolayer $WSe_2$. The latter is a non-conventional substrate that cannot be simply described by a model screening potential. The XAF-GW approach is easy to implement and opens the door to the use of GW methods on large interface systems, including other organic/2D material interfaces, and twisted bilayer systems.

## APPENDIX

This appendix provides the full derivation to show the validity of the Add-chi approximation in Cases 1 to 3 as considered in Section 2.2.

Case 1: The wavefunctions of the heterostructure (HS) are all completely localized on 1 or 2 only, and are given by the Kohn-Sham eigenfunctions of the individual components 1 or 2. The Kohn-Sham eigenspectrum of these states is the same as those of the individual components, but may be shifted by an additive constant.

Since components 1 and 2 are spatially separated, the matrix elements defined in Equation (4) are zero if the bra and ket wavevectors are localized on separate components. This immediately gives the result that

$$\chi^{HS}_{\vec{G}\vec{G}'}(\vec{q};0) = \chi^{1}_{\vec{G}\vec{G}'}(\vec{q};0) + \chi^{2}_{\vec{G}\vec{G}'}(\vec{q};0) \qquad (11)$$

We comment that if the energy differences $E_{v,\vec{k}+\vec{q}} - E_{c,\vec{k}}$ differ in the HS by $\delta$, the error is $O\left(\dfrac{\delta}{E_{v,\vec{k}+\vec{q}} - E_{c,\vec{k}}}\right)$.

Case 2: The wavefunctions of the HS are not completely localized on 1 or 2 only, but consist of a linear combination of the Kohn-Sham eigenfunctions of 1 and 2. Specifically, we assume that when 1 and 2 come into contact, the eigenvalues of 1 and 2 may shift relative to one another, but only by an additive constant (keeping the same energy differences between eigenvalues). When the eigenvalues associated with 1 and 2 are degenerate, the wavefunctions of the HS can be a linear combination of the corresponding eigenfunctions of 1 and 2. However, we assume that the eigenfunctions of 1 and 2 form an orthonormal set (in particular, eigenfunctions of 1 and 2 are assumed not to overlap in this case).

In our model, for a given $\vec{k}+\vec{q}$ (omitted in notation below), two valence wavefunctions from 1 and 2 mix to give a total of two distinct and orthonormal valence wavefunctions of the heterostructure:

$$|v_1^{HS}\rangle = \alpha_{11}|v_1\rangle + \alpha_{12}|v_2\rangle \tag{12}$$

$$|v_2^{HS}\rangle = \alpha_{21}|v_1\rangle + \alpha_{22}|v_2\rangle \tag{13}$$

This implies that the matrix $X = \begin{pmatrix} \alpha_{11} & \alpha_{12} \\ \alpha_{21} & \alpha_{22} \end{pmatrix}$ is a unitary matrix.

Similarly, for a given $\vec{k}$,

$$|c_1^{HS}\rangle = \beta_{11}|c_1\rangle + \beta_{12}|c_2\rangle \tag{14}$$

$$|c_2^{HS}\rangle = \beta_{21}|c_1\rangle + \beta_{22}|c_2\rangle \tag{15}$$

where $Y = \begin{pmatrix} \beta_{11} & \beta_{12} \\ \beta_{21} & \beta_{22} \end{pmatrix}$ is a unitary matrix.

For brevity of notation, for a given $\vec{q}$, we define the operator $g = e^{i(\vec{q}+\vec{G})\cdot\vec{r}}$ and $g' = e^{-i(\vec{q}+\vec{G'})\cdot\vec{r}}$ (dropping the index of $\vec{q}$). We further define $g_{ij} = \langle v_i|g|c_j\rangle$ and $g'_{ij} = \langle c_i|g'|v_j\rangle$. According to our assumption that the eigenfunctions of 1 and 2 do not overlap, $g_{ij}$ and $g'_{ij}$ will be zero for $i \neq j$.

Given $\vec{k}$ and $\vec{k}+\vec{q}$, we can write the contribution of $|v_1^{HS}\rangle$, $|v_2^{HS}\rangle$, $|c_1^{HS}\rangle$ and $|c_2^{HS}\rangle$ to $\chi_{\vec{G}\vec{G}'}^{HS}(\vec{q};0)$ as

$$\chi_{\vec{G}\vec{G}'}^{HS,partial}(\vec{q};0) = \frac{A+B+C+D}{E_{v,\vec{k}+\vec{q}} - E_{c,\vec{k}}} \tag{16}$$

where

$$A = \langle v_1^{HS}|g|c_1^{HS}\rangle\langle c_1^{HS}|g'|v_1^{HS}\rangle \tag{17}$$

$$B = \langle v_2^{HS}|g|c_2^{HS}\rangle\langle c_2^{HS}|g'|v_2^{HS}\rangle \tag{18}$$

$$C = \langle v_1^{HS}|g|c_2^{HS}\rangle\langle c_2^{HS}|g'|v_1^{HS}\rangle \tag{19}$$

$$D = \langle v_2^{HS}|g|c_1^{HS}\rangle\langle c_1^{HS}|g'|v_2^{HS}\rangle \tag{20}$$

Let $\gamma_1 = g_{11}g'_{11}$, $\gamma_2 = g_{22}g'_{22}$, $\gamma_3 = g_{22}g'_{11}$ and $\gamma_4 = g_{11}g'_{22}$. Then

$$A = |\alpha_{11}|^2|\beta_{11}|^2\gamma_1 + |\alpha_{12}|^2|\beta_{12}|^2\gamma_2 + \alpha_{12}^*\beta_{12}\beta_{11}^*\alpha_{11}\gamma_3 + \alpha_{11}^*\beta_{11}\beta_{12}^*\alpha_{12}\gamma_4 \tag{21}$$

$$B = |\alpha_{21}|^2|\beta_{21}|^2\gamma_1 + |\alpha_{22}|^2|\beta_{22}|^2\gamma_2 + \alpha_{22}^*\beta_{22}\beta_{21}^*\alpha_{21}\gamma_3 + \alpha_{21}^*\beta_{21}\beta_{22}^*\alpha_{22}\gamma_4 \tag{22}$$

$$C = |\alpha_{11}|^2|\beta_{21}|^2\gamma_1 + |\alpha_{12}|^2|\beta_{22}|^2\gamma_2 + \alpha_{12}^*\beta_{22}\beta_{21}^*\alpha_{11}\gamma_3 + \alpha_{11}^*\beta_{21}\beta_{22}^*\alpha_{12}\gamma_4 \tag{23}$$

$$D = |\alpha_{21}|^2|\beta_{11}|^2\gamma_1 + |\alpha_{22}|^2|\beta_{12}|^2\gamma_2 + \alpha_{22}^*\beta_{12}\beta_{11}^*\alpha_{21}\gamma_3 + \alpha_{21}^*\beta_{11}\beta_{12}^*\alpha_{22}\gamma_4 \tag{24}$$

Using the orthonormality of the HS wavefunctions and unitarity of X and Y, we can then obtain

$$A+B+C+D = \gamma_1 + \gamma_2 \tag{25}$$

which gives

$$\chi_{\vec{G}\vec{G}'}^{HS,partial}(\vec{q};0) = \chi_{\vec{G}\vec{G}'}^{1,partial}(\vec{q};0) + \chi_{\vec{G}\vec{G}'}^{2,partial}(\vec{q};0) \tag{26}$$

as required. The above analysis may be extended to all states involved in the summation for chi.

Case 3: We define a two-state hybridized model as follows. The valence band states of 1 and 2 overlap to form bonding and antibonding combinations in the HS, and likewise, the conduction band states of 1 and 2 overlap to form bonding and antibonding combinations in the HS. (We do not consider overlap between valence states of one component with conduction band states of the other component.) In the following, we keep up to first order all terms involving overlap matrix elements in the hybridization, and ignore second order terms.

We first work out the general solution for the HS Hamiltonian in the presence of hybridization between eigenfunctions of 1 and 2. Let $h_{HS}$, $h_1$ and $h_2$ be the Hamiltonians for the HS, components 1 and 2, respectively. Consider eigenfunctions $|\psi_1\rangle$ and $|\psi_2\rangle$ of $h_1$ and $h_2$ with the same eigenvalue $\varepsilon$. Let the eigenfunction of $h_{HS}$ be $|\psi\rangle = \alpha_1|\psi_1\rangle + \alpha_2|\psi_2\rangle$.

$$h_{HS}|\psi\rangle = \tilde{\varepsilon}|\psi\rangle \tag{27}$$

Assume $\langle\psi_1|h_{HS}|\psi_1\rangle = \langle\psi_2|h_{HS}|\psi_2\rangle = \varepsilon$, and let $\gamma = \langle\psi_1|h_{HS}|\psi_2\rangle$ and $s = \langle\psi_1|\psi_2\rangle$. This gives us

$$\begin{pmatrix} \varepsilon & \gamma \\ \gamma^* & \varepsilon \end{pmatrix}\begin{pmatrix} \alpha_1 \\ \alpha_2 \end{pmatrix} = \tilde{\varepsilon}\begin{pmatrix} 1 & s \\ s^* & 1 \end{pmatrix}\begin{pmatrix} \alpha_1 \\ \alpha_2 \end{pmatrix} \tag{28}$$

Solving equation (28), we obtain two solutions:

$$\psi_\pm = \frac{1}{\sqrt{2(1 \pm s)}}(\psi_1 \pm \psi_2) \tag{29}$$

with eigenvalues

$$\tilde{\varepsilon} = \frac{\varepsilon \pm \gamma}{1 \pm s} \approx \varepsilon(1 \mp s) \pm \gamma \tag{30}$$

In equations (29) and (30), we have written $\gamma$ and $s$ as real values, for simplicity and without loss of generality.

Since valence states of 1 and 2 can hybridize, and similarly, conduction states of 1 and 2 can hybridize, we define

$$\left|v_1^{HS}\right\rangle = \frac{1}{\sqrt{2(1+s_v)}}\left(\left|v_1\right\rangle + \left|v_2\right\rangle\right) \tag{31}$$

$$\left|v_2^{HS}\right\rangle = \frac{1}{\sqrt{2(1-s_v)}}\left(\left|v_1\right\rangle - \left|v_2\right\rangle\right) \tag{32}$$

$$\left|c_1^{HS}\right\rangle = \frac{1}{\sqrt{2(1+s_c)}}\left(\left|c_1\right\rangle + \left|c_2\right\rangle\right) \tag{33}$$

$$\left|c_2^{HS}\right\rangle = \frac{1}{\sqrt{2(1-s_c)}}\left(\left|c_1\right\rangle - \left|c_2\right\rangle\right) \tag{34}$$

We use again the notation in equations (16) to (20), and neglect terms involving $g_{ij}$ and $g'_{ij}$ for $i \neq j$, according to our model that includes only overlaps between valence states of the two components, or between conduction states of the two components.

The denominators in $\chi^{HS,partial}_{\vec{G}\vec{G}'}(\vec{q};0)$ corresponding to A, B, C and D are:

A:
$$\varepsilon_v - s_v\varepsilon_v + \gamma_v - \varepsilon_c + s_c\varepsilon_c - \gamma_c \equiv (\varepsilon_v - \varepsilon_c) - \Delta(\varepsilon_v - \varepsilon_c) \tag{35}$$

B:
$$\varepsilon_v + s_v\varepsilon_v - \gamma_v - \varepsilon_c - s_c\varepsilon_c + \gamma_c = (\varepsilon_v - \varepsilon_c) + \Delta(\varepsilon_v - \varepsilon_c) \tag{36}$$

C:
$$\varepsilon_v - s_v\varepsilon_v + \gamma_v - \varepsilon_c - s_c\varepsilon_c + \gamma_c \equiv (\varepsilon_v - \varepsilon_c) - \Delta_1(\varepsilon_v - \varepsilon_c) \tag{37}$$

D:
$$\varepsilon_v + s_v\varepsilon_v - \gamma_v - \varepsilon_c + s_c\varepsilon_c - \gamma_c = (\varepsilon_v - \varepsilon_c) + \Delta_1(\varepsilon_v - \varepsilon_c) \tag{38}$$

In the above, $\Delta$ and $\Delta_1$ are defined by equations (35) and (37), respectively.

Next, we define $\Delta' = s_v + s_c$ and $\Delta_1' = s_v - s_c$.

Using equations (21) to (24), and (31) to (38), we obtain, to first order in $s_v$, $s_c$,

$$\chi^{HS,partial}_{\vec{G}\vec{G}'}(\vec{q};0) = \frac{(\gamma_1+\gamma_2)}{4(\varepsilon_v-\varepsilon_c)}\left[\frac{1}{(1+\Delta')(1-\Delta)} + \frac{1}{(1-\Delta')(1+\Delta)} + \frac{1}{(1+\Delta_1')(1-\Delta_1)} + \frac{1}{(1-\Delta_1')(1+\Delta_1)}\right]$$

(39)

The first order terms in $\Delta$ vanish in equation (39), giving

$$\chi^{HS,partial}_{\vec{G}\vec{G}'}(\vec{q};0) = \chi^{1,partial}_{\vec{G}\vec{G}'}(\vec{q};0) + \chi^{2,partial}_{\vec{G}\vec{G}'}(\vec{q};0) \qquad (40)$$

as required. As for Case 2, the analysis can be applied to all states involved in the summation for chi.

## ACKNOWLEDGEMENTS

We acknowledge funding from Grant NRF-NRFF2013-07 from the National Research Foundation, Singapore, from Grant MOE2016-T2-2-132 from the Ministry of Education, Singapore, and support from the Singapore National Research Foundation, Prime Minister's Office, under its medium-sized centre program. Computations were performed on the NUS Graphene Research Centre cluster and National Supercomputing Centre Singapore (NSCC). We thank Miguel Dias Costa for systems support.

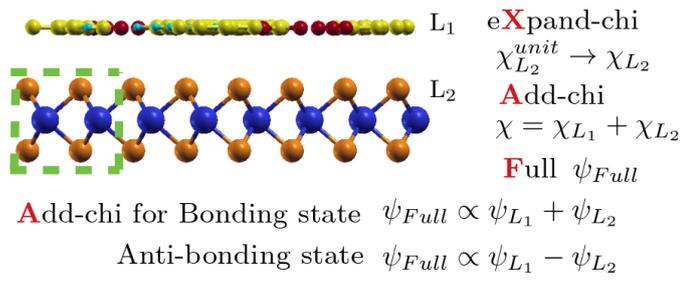

$L_1$   e**X**pand-chi
$\chi_{L_2}^{unit} \to \chi_{L_2}$

$L_2$   **A**dd-chi
$\chi = \chi_{L_1} + \chi_{L_2}$

**F**ull  $\psi_{Full}$

**A**dd-chi for Bonding state   $\psi_{Full} \propto \psi_{L_1} + \psi_{L_2}$

Anti-bonding state   $\psi_{Full} \propto \psi_{L_1} - \psi_{L_2}$

Supplementary Material for "Quasiparticle Levels at Large Interface Systems from Many-body Perturbation Theory: the XAF-GW method"

Fengyuan Xuan[1], Yifeng Chen[1] and Su Ying Quek[1,2,*]

[1]Centre for Advanced 2D Materials, National University of Singapore, Block S14, Level 6, 6 Science Drive 2, 117546, Singapore

[2]Department of Physics, National University of Singapore, 2 Science Drive 3, 117542, Singapore

* To whom correspondence should be addressed: phyqsy@nus.edu.sg

**In this supplementary material, we provide a detailed proof for Eq. (25) and (39) in the main text, as well as computational benchmarks to illustrate the reduction in computational resource possible with the XAF-GW method.**

*In the text, Eq. (25) is obtained by using the orthonormality of the HS wave functions and unitarity of X and Y to simplify Eq. (21)-(24). We provide a detailed proof of this step below.*

Consider the orthonormality of $|v_1>, |v_2>$ and $|v_1^{HS}>, |v_2^{HS}>$

$$\langle v_1|v_2\rangle = \langle v_1^{HS}|v_2^{HS}\rangle = 0$$

$$\langle v_1|v_1\rangle = \langle v_1^{HS}|v_1^{HS}\rangle = \langle v_2|v_2\rangle = \langle v_2^{HS}|v_2^{HS}\rangle = 1$$

So we have from the unitarity of $X$ ($XX^\dagger = I$):

$$\alpha_{11}^* \alpha_{21} + \alpha_{12}^* \alpha_{22} = 0$$

$$|\alpha_{11}|^2 + |\alpha_{12}|^2 = |\alpha_{21}|^2 + |\alpha_{22}|^2 = 1$$

Using $\alpha_{11}^* \alpha_{21} + \alpha_{12}^* \alpha_{22} = 0$, we have $\alpha_{11}^* = -\frac{\alpha_{12}^* \alpha_{22}}{\alpha_{21}}$. So we can show,

$$1 = |\alpha_{11}|^2 + |\alpha_{12}|^2 = \frac{|\alpha_{12}|^2|\alpha_{22}|^2}{|\alpha_{21}|^2} + |\alpha_{12}|^2 = \frac{|\alpha_{12}|^2}{|\alpha_{21}|^2}(|\alpha_{22}|^2 + |\alpha_{21}|^2) = \frac{|\alpha_{12}|^2}{|\alpha_{21}|^2}$$

$$|\alpha_{12}|^2 = |\alpha_{21}|^2$$

Thus we have

$$|\alpha_{11}|^2 + |\alpha_{21}|^2 = |\alpha_{12}|^2 + |\alpha_{22}|^2 = 1$$

We also have $\alpha_{11}^* = -\frac{\alpha_{12}^*\alpha_{22}}{\alpha_{21}} = -\frac{\alpha_{21}^*\alpha_{22}}{\alpha_{12}}$ from $X^\dagger X = I$

The same considerations hold for $\beta$.

Next, we first evaluate the coefficient of $\gamma_1$ in A+B+C+D.

$$|\alpha_{11}|^2|\beta_{11}|^2 + |\alpha_{21}|^2|\beta_{21}|^2 + |\alpha_{11}|^2|\beta_{21}|^2 + |\alpha_{21}|^2|\beta_{11}|^2$$
$$= (|\alpha_{11}|^2 + |\alpha_{21}|^2)|\beta_{11}|^2 + (|\alpha_{21}|^2 + |\alpha_{11}|^2)|\beta_{21}|^2 = |\beta_{11}|^2 + |\beta_{21}|^2 = 1$$

Similarly the coefficient of $\gamma_2$ in A+B+C+D is also unity. Then we show the coefficient of $\gamma_3$ in A+B+C+D vanishes.

$$\alpha_{12}^*\beta_{12}\beta_{11}^*\alpha_{11} + \alpha_{22}^*\beta_{22}\beta_{21}^*\alpha_{21} + \alpha_{12}^*\beta_{22}\beta_{21}^*\alpha_{11} + \alpha_{22}^*\beta_{12}\beta_{11}^*\alpha_{21}$$
$$= (\alpha_{12}^*\alpha_{11} + \alpha_{22}^*\alpha_{21})\beta_{12}\beta_{11}^* + (\alpha_{22}^*\alpha_{21} + \alpha_{12}^*\alpha_{11})\beta_{22}\beta_{21}^* = 0$$

where we use the equation shown above $\alpha_{11}^* = -\frac{\alpha_{21}^*\alpha_{22}}{\alpha_{12}}$, we have $\alpha_{22}^*\alpha_{21} + \alpha_{12}^*\alpha_{11} = \alpha_{11}^*\alpha_{12} + \alpha_{21}^*\alpha_{22} = 0$. Finally we show the coefficient of $\gamma_4$ in A+B+C+D vanishes.

$$\alpha_{11}^*\beta_{11}\beta_{12}^*\alpha_{12} + \alpha_{21}^*\beta_{21}\beta_{22}^*\alpha_{22} + \alpha_{11}^*\beta_{21}\beta_{22}^*\alpha_{12} + \alpha_{21}^*\beta_{11}\beta_{12}^*\alpha_{22}$$
$$= (\alpha_{11}^*\alpha_{12} + \alpha_{21}^*\alpha_{22})\beta_{11}\beta_{12}^* + (\alpha_{21}^*\alpha_{22} + \alpha_{11}^*\alpha_{12})\beta_{21}\beta_{22}^* = 0$$

Thus we show how Eq. (25) is derived.

*Detailed proof for Eq. (39)-(40):*

Inserting Eq. (30) – (34) to $\chi_{GG'}^{HS,partial}$, and using the result in Case 2, we obtain,

$$\chi_{GG'}^{HS,partial} = \frac{\langle v_1^{HS}|g|c_1^{HS}\rangle\langle c_1^{HS}|g'|v_1^{HS}\rangle}{\tilde{\varepsilon}_{v1} - \tilde{\varepsilon}_{c1}} + \frac{\langle v_2^{HS}|g|c_2^{HS}\rangle\langle c_2^{HS}|g'|v_2^{HS}\rangle}{\tilde{\varepsilon}_{v2} - \tilde{\varepsilon}_{c2}}$$
$$+ \frac{\langle v_1^{HS}|g|c_2^{HS}\rangle\langle c_2^{HS}|g'|v_1^{HS}\rangle}{\tilde{\varepsilon}_{v1} - \tilde{\varepsilon}_{c2}} + \frac{\langle v_2^{HS}|g|c_1^{HS}\rangle\langle c_1^{HS}|g'|v_2^{HS}\rangle}{\tilde{\varepsilon}_{v2} - \tilde{\varepsilon}_{c1}}$$
$$= \frac{1}{\varepsilon_v - s_v\varepsilon_v + \gamma_v - \varepsilon_c + s_c\varepsilon_c - \gamma_c} \frac{\gamma_1 + \gamma_2}{4(1+s_v)(1+s_c)}$$
$$+ \frac{1}{\varepsilon_v + s_v\varepsilon_v - \gamma_v - \varepsilon_c - s_c\varepsilon_c + \gamma_c} \frac{\gamma_1 + \gamma_2}{4(1-s_v)(1-s_c)}$$
$$+ \frac{1}{\varepsilon_v - s_v\varepsilon_v + \gamma_v - \varepsilon_c - s_c\varepsilon_c + \gamma_c} \frac{\gamma_1 + \gamma_2}{4(1+s_v)(1-s_c)}$$
$$+ \frac{1}{\varepsilon_v + s_v\varepsilon_v - \gamma_v - \varepsilon_c + s_c\varepsilon_c - \gamma_c} \frac{\gamma_1 + \gamma_2}{4(1-s_v)(1+s_c)}$$

The first term can be simplified as,

$$\frac{1}{\varepsilon_v - s_v\varepsilon_v + \gamma_v - \varepsilon_c + s_c\varepsilon_c - \gamma_c} \frac{\gamma_1 + \gamma_2}{4(1+s_v)(1+s_c)}$$
$$= \frac{1}{\varepsilon_v - \varepsilon_c - s_v\varepsilon_v + \gamma_v + s_c\varepsilon_c - \gamma_c} \frac{\gamma_1 + \gamma_2}{4(1+s_v)(1+s_c)}$$
$$= \frac{1}{4(\varepsilon_v - \varepsilon_c)} \frac{1}{1 - (s_v\varepsilon_v - \gamma_v - s_c\varepsilon_c + \gamma_c)/(\varepsilon_v - \varepsilon_c)} \frac{\gamma_1 + \gamma_2}{(1+s_v)(1+s_c)}$$
$$= \frac{1}{4(\varepsilon_v - \varepsilon_c)} \frac{1}{1 - \Delta} \frac{\gamma_1 + \gamma_2}{(1+s_v)(1+s_c)}$$

where we define $\Delta= (s_v\varepsilon_v - \gamma_v - s_c\varepsilon_c + \gamma_c)/(\varepsilon_v - \varepsilon_c)$. We can do the same factorization for the second term. The third and fourth terms can be similarly simplified by defining $\Delta_1= (s_v\varepsilon_v - \gamma_v + s_c\varepsilon_c - \gamma_c)/(\varepsilon_v - \varepsilon_c)$. Given that

$$\frac{1}{(1+s_v)(1+s_c)} \approx \frac{1}{1+s_v+s_c} = \frac{1}{1+\Delta'}$$

$$\frac{1}{(1+s_v)(1-s_c)} \approx \frac{1}{1+s_v-s_c} = \frac{1}{1+\Delta'_1}$$

where $\Delta'= s_v + s_c$ and $\Delta'_1= s_v - s_c$, and keeping first order terms in $s_v$ and $s_c$, we get

$$\begin{aligned}
\chi_{GG'}^{HS,partial} &= \frac{1}{4(\varepsilon_v - \varepsilon_c)}\bigg[\frac{1}{1-\Delta}\frac{\gamma_1+\gamma_2}{(1+s_v)(1+s_c)} + \frac{1}{1+\Delta}\frac{\gamma_1+\gamma_2}{(1-s_v)(1-s_c)} \\
&\quad + \frac{1}{1-\Delta_1}\frac{\gamma_1+\gamma_2}{(1+s_v)(1-s_c)} + \frac{1}{1+\Delta_1}\frac{\gamma_1+\gamma_2}{(1-s_v)(1+s_c)}\bigg] \\
&\approx \frac{\gamma_1+\gamma_2}{4(\varepsilon_v - \varepsilon_c)}\bigg[\frac{1}{1-\Delta}\frac{1}{1+\Delta'} + \frac{1}{1+\Delta}\frac{1}{1-\Delta'} + \frac{1}{1-\Delta_1}\frac{1}{1+\Delta'_1} + \frac{1}{1+\Delta_1}\frac{1}{1-\Delta'_1}\bigg] \\
&\approx \frac{\gamma_1+\gamma_2}{4(\varepsilon_v - \varepsilon_c)}[1 - \Delta' + \Delta + 1 + \Delta' - \Delta + 1 - \Delta'_1 + \Delta_1 + 1 + \Delta'_1 - \Delta_1] \\
&= \frac{\gamma_1+\gamma_2}{(\varepsilon_v - \varepsilon_c)} = \chi_{GG'}^{1,partial} + \chi_{GG'}^{2,partial}
\end{aligned}$$

where we have kept all terms up to first order in $s_v$ and $s_c$ (using the fact that $\Delta$, $\Delta'$, $\Delta_1$, $\Delta'_1$ are all first order in $s_v$ and $s_c$). Thus we show that the first order term in $s_v$ and $s_c$ vanishes and Eq. (40) is obtained.